\documentclass[10pt,conference]{IEEEtran}
\IEEEoverridecommandlockouts
\usepackage{cite}
\usepackage{amsmath,amssymb,amsfonts}
\usepackage{algorithmic}
\usepackage{graphicx}
\usepackage{textcomp}
\usepackage{xcolor}
\usepackage{stfloats}
\usepackage{subcaption}
\def\BibTeX{{\rm B\kern-.05em{\sc i\kern-.025em b}\kern-.08em
    T\kern-.1667em\lower.7ex\hbox{E}\kern-.125emX}}

\usepackage{tabularx}

\usepackage[nolist]{acronym}
\newacro{SUMO}{Simulation of Urban Mobility}
\newacro{RT}{Ray-Tracing}
\newacro{5G}{5th generation}
\newacro{6G}{6th generation}
\newacro{B5G}{beyond 5G}
\newacro{3D}{3th dimensional}
\newacro{OSM}{OpenStreetMap}
\newacro{API}{Application Programming Interface}
\newacro{CSV}{Comma-separated values}
\newacro{4D}{4th dimensional}
\newacro{LOS}{line of sight}
\newacro{Tx}{transmitter}
\newacro{Rx}{receiver}
\newacro{WI}{Wireless InSite}
\newacro{MIMO}{Multiple-Input Multiple-Output}
\newacro{NMSE}{Normalized Mean Square Error}
\newacro{ULA}{Uniform Linear Array}
\newacro{LSTM}{Long Short-Term Memory}
\newacro{BT}{Beam-Tracking}
\newacro{BS}{Base Station}
\newacro{UE}{User Equipament}
\newacro{CT}{Channel Tracking}
\newacro{RIS}{Reconfigurable Intelligent Surface}
\newacro{CSI}{Channel State Information}
\newacro{KF}{Kalman Filter}
\newacro{RLS}{Recursive Least Squares}
\newacro{GPR}{Gaussian Process Regression}
\newacro{mmWave}{millimeter Wave}
\newacro{LS}{Least Squares}
\newacro{ML}{Machine Learning}
\newacro{SNR}{Signal to Noise Ratio}
\newacro{NN}{Neural Network}
\newacro{CNN}{Convolutional Neural Network}
\newacro{GNSS}{Global Navigation Satellite System}
\newacro{AR}{Auto-regressive}

\begin{document}

\title{Machine Learning-Based mmWave MIMO Beam Tracking in V2I Scenarios: Algorithms and Datasets\\
\thanks{This work was supported by RNP, with resources from MCTIC, Grant \#01245.010645/2020-14, under the ’Brazil 6G Project’. This work was also supported in part by the Innovation Center, Ericsson Telecomunica\c{c}\~{o}es S.A., Brazil, CNPq/Capes, Brazil,
and
by the Project Smart 5G Core And MUltiRAn Integration (SAMURAI) (MCTIC/CGI.br/FAPESP Grant 2020/05127-2).}
}
\author{\IEEEauthorblockN{
    Ailton Oliveira\IEEEauthorrefmark{1},
    Daniel Suzuki\IEEEauthorrefmark{1}, 
    Sávio Bastos\IEEEauthorrefmark{1},
    Ilan Correa\IEEEauthorrefmark{1},
    Aldebaro Klautau\IEEEauthorrefmark{1}}
    \IEEEauthorblockA{
        \IEEEauthorrefmark{1} LASSE - Telecommunications, Automation and Electronics Research and Development Center, Belém-PA, Brazil   \\
        Email: \IEEEauthorrefmark{1}\{ailton.pinto, daniel.suzuki, savio.bastos\}@itec.ufpa.br, \{ilan, aldebaro\}@ufpa.br}
}

\maketitle

\begin{abstract}
This work investigates the use of machine learning applied to the beam tracking problem in 5G networks and beyond. The goal is to decrease the overhead associated to MIMO millimeter wave  beamforming. In comparison to beam selection (also called initial beam acquisition), ML-based beam tracking is less investigated in the literature due to factors such as the lack of comprehensive datasets. One of the contributions of this work is a new public multimodal dataset, which includes images, LIDAR information and GNSS positioning, enabling the evaluation of new data fusion algorithms applied to wireless communications. The work also contributes with an evaluation of the performance of beam tracking algorithms, and associated methodology. When considering as inputs the LIDAR data, the coordinates and the information from previously selected beams, the  proposed  deep neural network based on ResNet and using LSTM layers, significantly outperformed the other beam tracking models.
\end{abstract}

\begin{IEEEkeywords}
Beam Tracking, mmWave, deep learning, dataset, V2I.
\end{IEEEkeywords}

\section{Introduction}
\label{sec:introduction}


The current fifth generation (5G) and the incoming sixth generation wireless communications envisages, among other features, very high peak data rate and very low latency. One way of fulfilling the data rate and latency requirements is with the use of higher bandwidths. However, due to the scarcity spectrum at the currently mostly used sub-6~GHz frequencies, wider bandwidths are being reserved for mobile communications at the millimeter wave (mmWave) band, such as 28~GHz and 60~GHz~\cite{Roh14}. With greater amounts of available spectrum at mmWave, decreasing symbol time, decreasing latency, and increasing throughput are possible. A drawback of the mmWave bands is the higher attenuation compared to sub-6~GHz frequencies. 

Because of the higher attenuation of the mmWave bands, multiple input multiple output (MIMO) techniques are one of the main technologies of the 5G development at mmWave bands \cite{mmWave2020}, especially beamforming techniques. Beamforming provides better directionality of the electromagnetic wave, allowing it to circumvent the high path attenuation~\cite{Bjornson19}.A main challenge of massive MIMO at mmWave frequencies is that the beams formed by the beamforming techniques can be very narrow, requiring that the beams of the transmitter and the receiver point to each other. 

In this context,  
exploring all possibilities to optimize the direction of transmission beams between \ac{BS} and \ac{UE} using ``brute force'' beam sweeping leads to significant overhead.
There are several algorithms to reduce this overhead.
For example, a process called \emph{beam training} is part of standards such IEEE~802.11ad, 5G, and other wireless networks~\cite{kim_fast_2014,zhou_enhanced_2017}. When the number of antennas increases due to the high number of possible beams, the search for the best beam index can be a time-consuming process~\cite{heath_overview_2016}. 

 
As a mean to solve this issue, many recent works are applying \ac{ML} techniques to predict the beams~\cite{Wenyan20,Yang19}. However, this work relies on exchanged pilot signals to allow channel estimation and the prediction of the beam index~\cite{Qiang21}, which is also challenging due to a high number of antennas, which require a high number of pilots to be exchanged, consuming wireless resources. 


Other studies have investigated beam tracking as a potential solution to the overhead issue. In \cite{Marasinghe2022}, the authors employed LIDAR in an indoor scenario with mobile users, achieving a prediction accuracy of 88.7\%. Similarly, \cite{Rinchi2023} utilized LIDAR data, incorporating real measured data as input to a long short-term memory (LSTM) network to predict future beam indices, resulting in an accuracy of 84\%. This study modeled the LIDAR input based on the angle and distance to each obstacle. The research presented in \cite{jiang2022lidar} used the same input parameters as the study mentioned earlier. Their proposed architecture leveraged all LIDAR inputs from the preceding instance, with performance compared against a baseline that predicted using prior beam knowledge. This approach yielded an accuracy of 57.5\% for current beam prediction, which was close to the baseline with the previous beam information.


In this paper, we evaluate the prediction of the beam index using information available for only one of the devices on the communication pair to avoid exchanging pilots. Thus, the technique evaluated in this work is an alternative to overcome the issues of finding the optimum beam index and the exchange of pilots previously discussed. The context is a mobile network with vehicles connected and has a variety of sensors. The out-of-band data from the sensors could be used to find the directions, i.e., the beam index, based on the vehicle's surroundings. In this context, we evaluate the feasibility of using \ac{GNSS} and LIDAR data combined with previous beam directions to find the current best beam index. In summary, the contributions of this work can be organized as follows:
\begin{itemize}
    \item A Deep Learning model that uses spatial data and previous beam indices to perform an estimation of the next best beam.
    
    \item New datasets following Raymobtime methodology \cite{Klautau2018} for beam tracking applications with time correlation.
\end{itemize}

This paper is organized as follows. Section~\ref{sec:system_model} presents the system model, which also discusses finding the optimum beam pair for the receiver and the transmitter and its influence on communication performance. Section~\ref{sec:dataset} describes the dataset used to generate the results presented in this work. Section~\ref{sec:evaluations} presents details of the evaluation setup, which includes the adopted \ac{ML} model and the communication system, and also presents the results of our evaluations. Section~\ref{sec:conclusions} presents the conclusions of the paper.




\section{System Model}
\label{sec:system_model}
\subsection{Communication System Model}
In this paper, we consider the downlink of a mobile communication system, which operates at \ac{mmWave} frequencies and uses a fully digital MIMO architecture. The \ac{BS} and the \ac{UE} have an \ac{ULA} with $N$ antenna elements, $N_{tx}$ for the transmitter and $N_{rx}$ for the receiver, spaced by half wavelength ($\lambda/2$).

The channel matrix $\mathbf{H} \in C^{N_{tx} \times N_{rx}}$ is obtained from multipath parameters obtained with ray-tracing and modeled according to the geometric channel model \cite{Heath16} for the narrowband case as
\begin{align}
    \label{eq:nb_H}
    \mathbf{H} = \sum_{l=0}^{L-1} \alpha_l \mathbf{a}_r(\phi_l^A,\theta_l^A) \mathbf{a}_t^H(\phi_l^D,\theta_l^D).
\end{align}
where $L$ is the number of rays, and $\alpha_l$ is the gain for the $l$-th ray. For the \ac{ULA} case, it uses the steering vectors $\textbf{a}(\phi_l, \lambda_l) = \textbf{a}(\phi_l) \otimes \textbf{a}(\theta_l)$ that can be described as
\begin{align}
    \textbf{a}(\phi_l) = [1, e^{-i2\pi\sin(\phi_l)d\lambda}, ..., e^{-i2\pi\sin(\phi_l)d\lambda(N-1)}],
\end{align}
The evolution of channel in time can be simulated using different strategies. One that is widely used due to its simplicity is the adoption of a first-order  \ac{AR} model~\cite{RLS_SGD}. In contrast, this work adopts a more realistic approach for simulating the channel evolution, which relies on the Raymobtime methodology \cite{klautau2021generating}. Raymobtime simulates the channel evolution, keeping consistency over time, frequency, and space~\cite{klautau2021generating}.  

\begin{figure}[h]
    \centering
    \includegraphics[width=8cm]{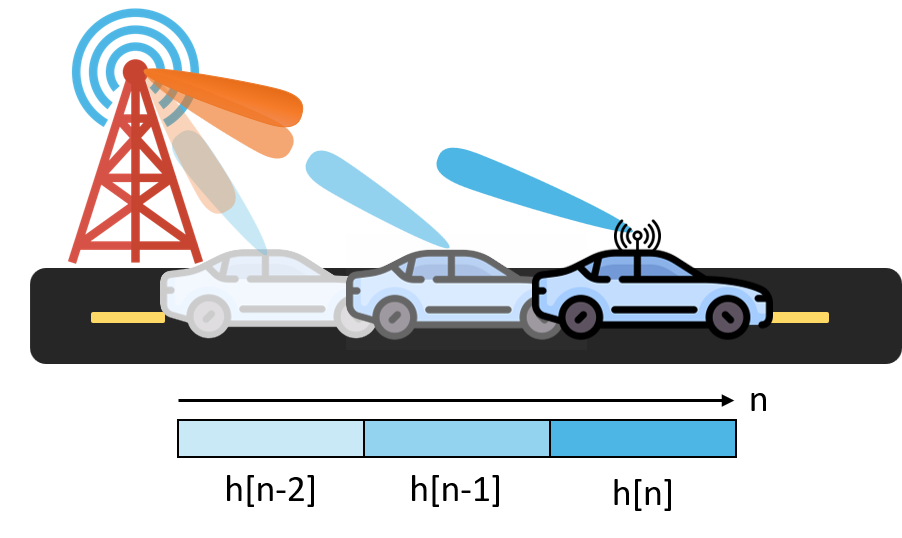}
    \caption{Beam sweeping over discrete-time $n$ in a V2I scenario.}
    \label{fig:beamforming}
\end{figure}

\subsection{Beamforming}


Effective beamforming relies on accurate \ac{CSI} to dynamically adjust the beam patterns in real-time, as shown in Fig.~\ref{fig:beamforming}, ensuring optimal performance even in the presence of user mobility and varying environmental conditions.

In this paper, we assume that both the \ac{BS} and \ac{UE} employ beamforming, such that the received signal at the UE is 
\begin{equation}
    y_i =\mathbf{w}_i^{*}\mathbf{H} \mathbf{f}_i,
\end{equation}
where $\mathbf{w}_i$ and $\mathbf{f}_i$ are respectively the precoding and combiner vectors used at the BS and UE to perform the beamforming. The vectors $\mathbf{w}_p$ and $\mathbf{f}_q$ are chosen from the codebooks $\mathcal{C}_t = \{\mathbf{w}_1,~\text{...},~\mathbf{w}_{|\mathcal{C}_t|}\}$ and $\mathcal{C}_r = \{\mathbf{f}_1,~\text{...},~\mathbf{f}_{|\mathcal{C}_r|}\}$, where $|\mathcal{C}_t|$ and $|\mathcal{C}_r|$ are drawn from the Discrete Fourier Transform (DFT)~\cite{he2017codebook} as
\begin{equation}
  \mathbf{F}(n) = \frac{1}{\sqrt{M}} e^{-\frac{j2\pi(n-1)}{M}}, 
\end{equation}
where $M$ is the number of codewords in the codebook and $n = 1,~\text{...},~N$.

Thus, the \emph{optimal} beam index $\hat i$ is given by

\begin{align}
\hat i = \arg\max_{i\in\{1, \cdots, M\}} |y_i|.
\end{align}

\subsubsection{Beam Selection}

Beam selection refers to the process of choosing the best beam from a predefined set of beams (codebook) that maximizes the received signal strength. This can be achieved by evaluating the received signal $|y_i|$ for each beam $i$ and selecting the beam index $\hat i$ that provides the highest signal strength. When the beams are ``pointy'' (e.\,g., as when using a DFT codebook~\cite{heath_overview_2016}), the goal is to ensure that the selected beam aligns with the strongest path of the transmitted signal.

\subsubsection{Beam Tracking}

Beam tracking involves continuously updating the selected ``pointy'' beam to maintain the alignment with the strongest signal path as the user moves or as the environment changes. This requires periodic measurements and adjustments to the beam index $\hat i$ to adapt to the dynamic nature of the wireless channel. Beam tracking is essential for maintaining optimal communication performance in mobile scenarios and in environments with high variability.

\section{Datasets and Methodology}
\label{sec:dataset}

\begin{figure}[ht]
    \centering
    \begin{subfigure}[b]{0.45\textwidth}
        \centering
        \caption{}
        \includegraphics[width=0.8\textwidth]{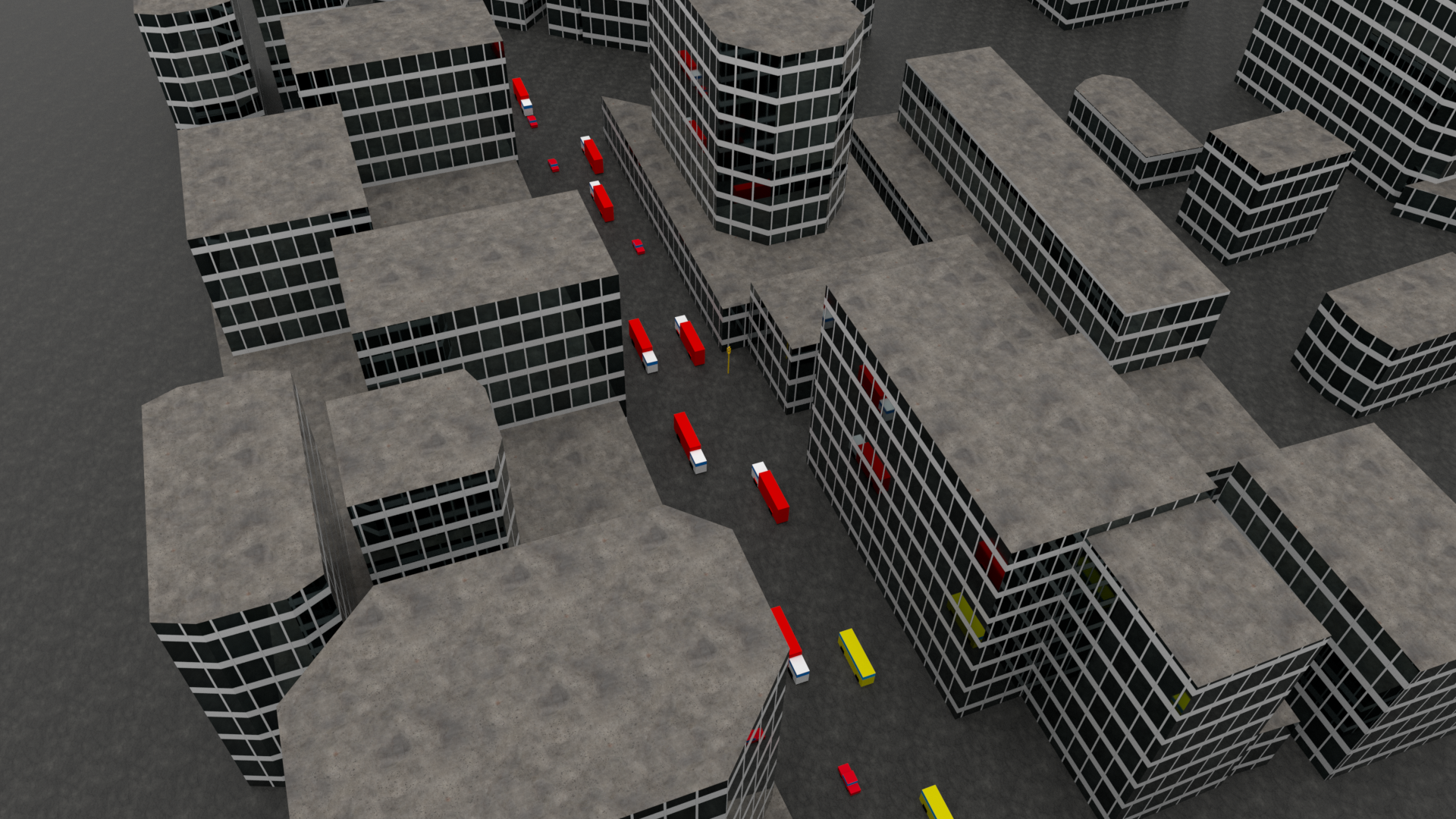}
        \label{fig:rosslyn}
    \end{subfigure}
    \hfill
    \begin{subfigure}[b]{0.45\textwidth}
        \centering
        \caption{}
        \includegraphics[width=0.8\textwidth]{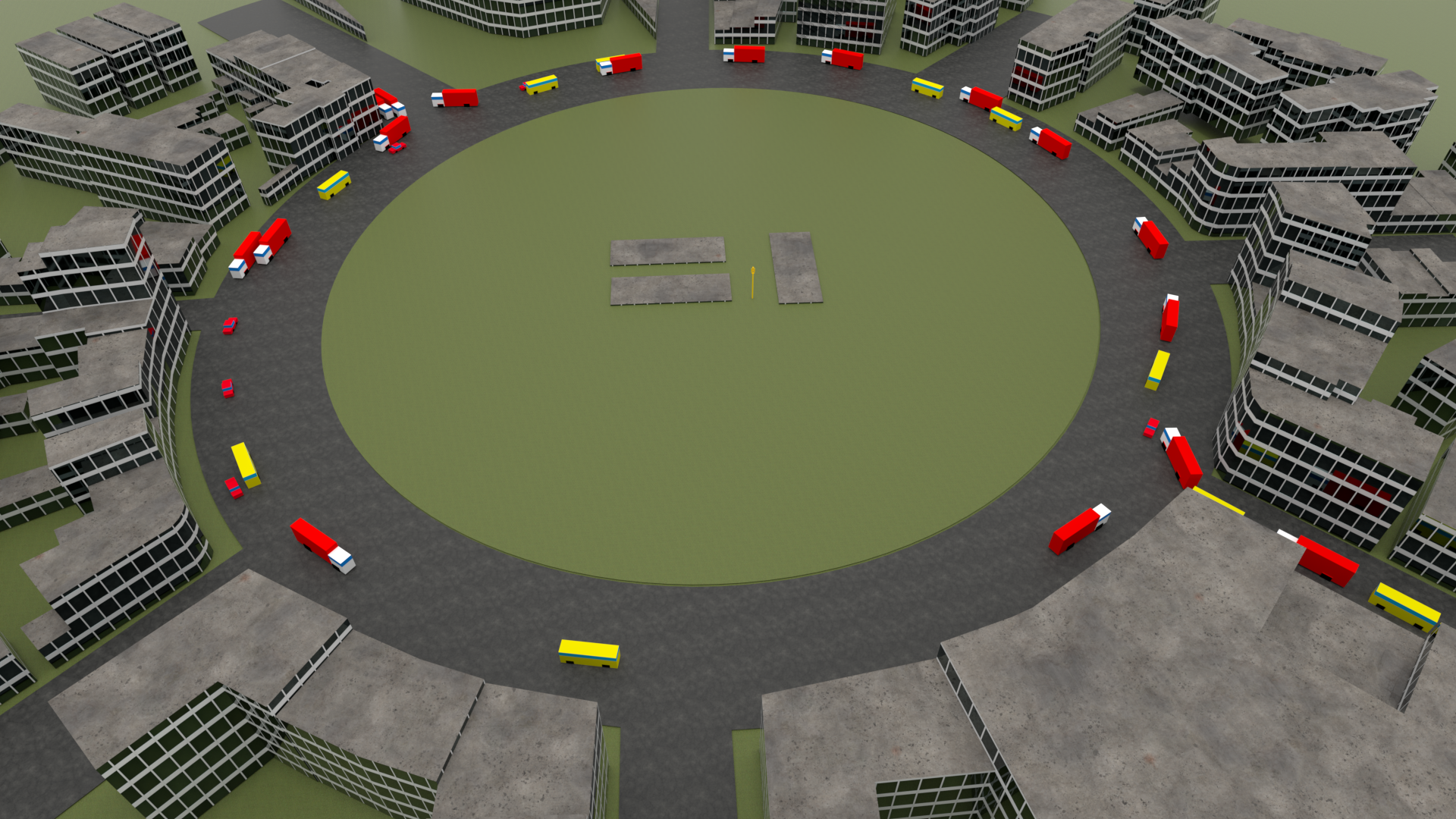}
        \label{fig:porto}
    \end{subfigure}
    \caption{Simulated scenarios: (a) Rosslyn (in USA) urban-canyon and (b) Porto (in Portugal) roundabout.}
    \label{fig:main}
\end{figure}

The Raymobtime methodology, initially proposed in~\cite{Klautau2018} and later refined in~\cite{klautau2021generating}, enables generating realistic wireless channel data with time evolution in simulation scenarios based on real locations, as shown in Fig~\ref{fig:main}, and data extracted from public sources~\cite{oliveira2020ray}.
The datasets created using this methodology are particularly designed for \ac{ML} applications, focusing on vehicle-to-infrastructure (V2I) and vehicle-to-vehicle (V2V) communications, thereby providing rich resources for advancing research in these areas.

Raymobtime datasets are organized into episodes and scenes, providing a structured way to capture the dynamics of wireless communication environments over time. An episode represents a complete set of data for a given simulation run, capturing the entire process from beginning to end under specific conditions (e.g., environment, frequency, mobility). Each episode consists of multiple scenes that chronologically represent the progression of the simulation. By examining changes across scenes within an episode, researchers can study how factors like mobility, environmental changes, and varying signal conditions impact communication performance.

\begin{table*}[t]
\caption{Raymobtime datasets specifically designed for beam tracking.}
\begin{center}
\begin{tabular}{|c|c|c|c|c|c|c|c|}
\hline
\multicolumn{1}{|c|}{Dataset} &
\multicolumn{1}{|c|}{Frequency} &
\multicolumn{1}{c|}{Location} &
\multicolumn{1}{c|}{Scenario} &
\multicolumn{1}{c|}{Receivers} &
\multicolumn{1}{c|}{\begin{tabular}[c]{@{}c@{}}Scenes\\ per episode\end{tabular}} &
\multicolumn{1}{c|}{Episodes} &
\multicolumn{1}{c|}{Time between scenes} \\
\hline
    t001 & 28GHz & Rosslyn & Urban-Canyon & 2 & 10 & 5000 & 20ms \\
    \hline
    t002 & 28GHz & Porto & Roundabout & 5 & 20 & 250 & 20ms\\
    \hline
\end{tabular}
\end{center}
\label{tab:tracking datasets}
\end{table*}

\subsection{Multimodal data}
\begin{figure}[h]
    \centering
    \includegraphics[width=0.36\textwidth]{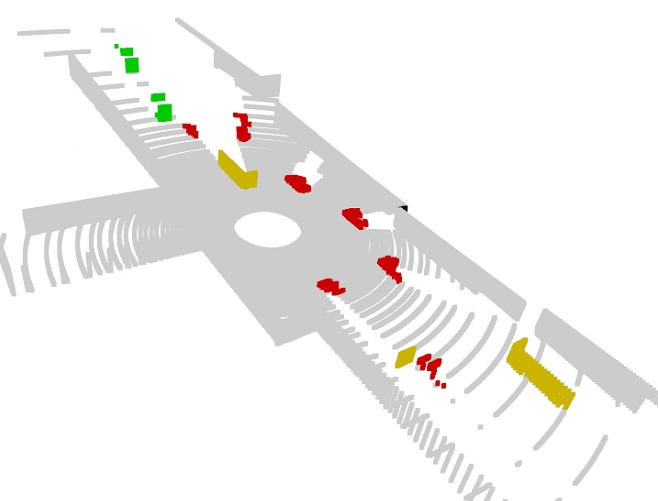}
    \caption{Point cloud obtained from a LIDAR placed in a vehicle.}
    \label{fig:lidar}
\end{figure}

As mentioned, the Raymobtime dataset offers a comprehensive suite of data types to facilitate different approaches to optimize and solve problems in 5G/6G networks. More specifically, some key features of the dataset include:

\begin{itemize}
    \item \textbf{Wireless Channel Data:}
    \begin{itemize}
        \item Ray-tracing simulations data of the wireless channel, capturing the multipath propagation characteristics and signal interactions with the environment. It includes information such as gain, phase, departure and arrival angles, time of arrival, and ray's interactions.
    \end{itemize}
    \vspace{1cm}
    \item \textbf{LIDAR Data:}
    \begin{itemize}
        \item High-resolution LIDAR point cloud data capturing the 3D structure and scatters at the environment. The sensor can be placed at the \ac{BS} or at the \ac{UE}, as shown in Fig.~\ref{fig:lidar}.
    \end{itemize}

    \item \textbf{GNSS Data:}
    \begin{itemize}
        \item Geographical localization of objects in the scenario, including the transmitter and receiver, enabling the correlation of spatial positioning with channel characteristics.
    \end{itemize}

    \item \textbf{Image Data:}
    \begin{itemize}
        \item Visual context of the simulation environment through images. This data aids in visual verification and can be used in conjunction with LIDAR and GNSS data for comprehensive environmental modeling.
    \end{itemize}
\end{itemize}

The combination of these diverse data types provides a rich resource for developing and testing \ac{ML} algorithms.

\subsection{Tracking scenarios}
The \ac{BS} can serve a variety of devices, including connected vehicles, which in general move faster than other \ac{UE}, thereby posing more strict requirements on the network. To effectively explore and evaluate beam tracking performance, two datasets in different scenarios were made to address these requirements.

The first dataset is the Rosslyn scenario (Fig.~\ref{fig:rosslyn}), which is a simulation environment validated for diverse papers~\cite{zheng2021deep,reus2021deep, mashhadi2021federated}, using two \ac{BS} one in each sidewalk in order to create a more balanced dataset. This dataset provides insights into how beam tracking algorithms perform in challenging urban landscapes.

The second dataset focuses on scenarios involving vehicles navigating a busy roundabout (Fig.~\ref{fig:porto}). Vehicles in this dataset travel at varying speeds up to 60 km/h, while changing their orientation vector due to the curved trajectories. This dataset helps evaluate the adaptability and precision of beam tracking methods in complex, high-mobility intersections.

Both datasets were created with episodes with a larger number of scenes and a time between scenes of 20ms, which was considered an adequate coherence time for the beams. Table~\ref{tab:tracking datasets} summarizes both datasets' key parameters and characteristics.

\section{Beam tracking Deep Learning model}
\label{sec: Algorithms}


\begin{figure*}[t]
    \centering
    \includegraphics[width=0.95\textwidth]{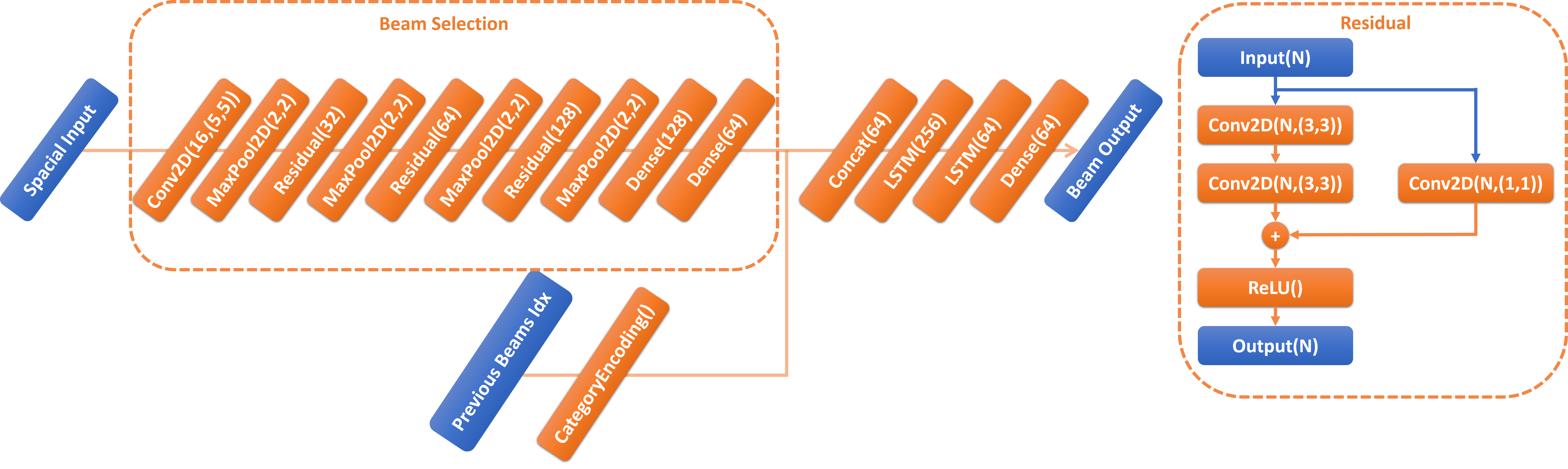}
    \caption{Proposed beam tracking hybrid architecture.}
    \label{fig:Architecture}
\end{figure*}

As shown in Fig.~\ref{fig:Architecture}, the proposed tracking method is realized by a hybrid architecture joining a beam-selection \ac{CNN} based on the ResNet framework \cite{he2016deep}, designed to extract high-level spatial features from the input data, and a \ac{LSTM} model designed to handle both the \ac{CNN} output and the historical beam indices.

The \ac{CNN} model  is capable of processing two types of input matrices:

\begin{figure}[bht]
    \centering
    \includegraphics[width=0.5\textwidth]{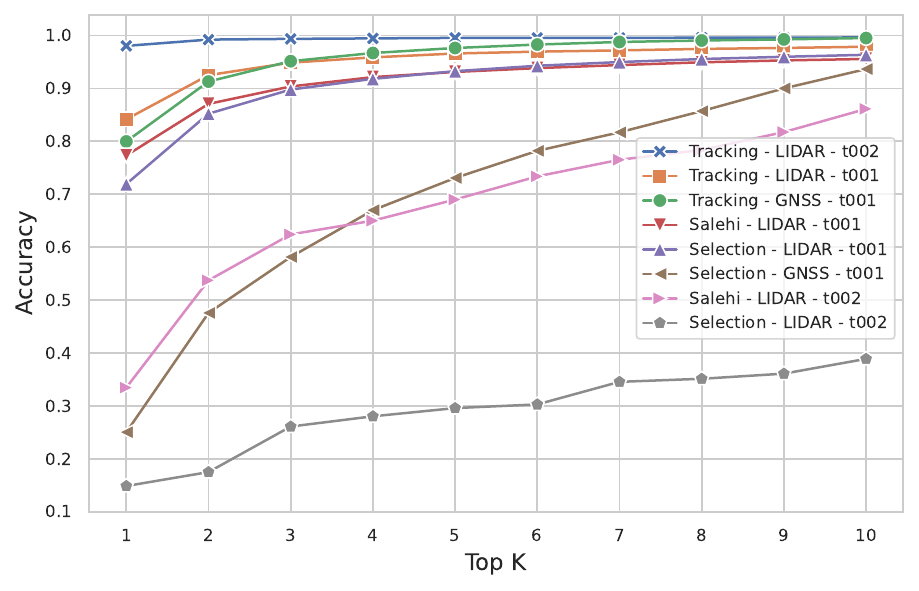}
    \caption{Top-K accuracy for each architecture with GNSS and LIDAR.}
    \label{fig:topk}
\end{figure}

\begin{itemize}
    \item LIDAR Matrix: This matrix represents the spatial information captured by LIDAR sensors. The raw data is quantized and transformed into a matrix, where each position functions similarly to a voxel: A value of -1 indicates an obstacle, 0 indicates no scatterers, -2 indicates the \ac{BS} position, and -3 indicates the \ac{UE} position. This processing method follows the approach described in \cite{dias2019position}.
    \item Coordinate Matrix based on GNSS: This matrix contains positional data from all objects in the scene, as outlined in \cite{suzuki2022ray}. For every scatter (building and cars that are not Tx or Rx) has a value of 1, the transmitter has a value of 10, and the receiver has a gradient of values that start at 3 and grow in the direction in which the vehicle is moving.
\end{itemize}

The LSTM model processes a concatenated input consisting of the \ac{CNN} output and an observation window of the previous $W_b$ beam indices. This design allows the LSTM to capture temporal dependencies and trends in the beam selection process, enhancing the overall tracking performance. For the experimental evaluation, all results consider a $W_b$  = $3$ and $64$ beams in a ULA array.

\section{Experimental Evaluation}
\label{sec:evaluations}

To evaluate the proposed architecture, the models were trained and tested in both scenarios independently, using LIDAR and coordinate matrix as inputs, with the test results shown in Fig. \ref{fig:topk}. 

To assess the performance of the proposed hybrid \ac{NN} for tracking, we compared the top-K accuracy with the beam selection component of the \ac{NN} and Salehi's LIDAR beam selection \ac{NN} from \cite{salehi2022deep}. For scenario t001, we compared the tracking, selection, and Salehi architectures using both LIDAR and coordinate (GNSS) data. In the second scenario, the same architectures were tested using only LIDAR data.


For the t001 scenario, the results, presented in Fig. \ref{fig:topk}, show that both the selection achieves results similar to other works in the literature~\cite{zheng2021deep,morais2023position}, with our beam selection achieving 59\% of accuracy in top-1 using LIDAR, and Salehi's \ac{NN} achieved 78\% accuracy in top-1. However, The hybrid tracking architecture achieved superior performance compared to the other baselines, with the best performance being 84\% accuracy in top-1 using LIDAR data in the t001 scenario.

In the t002 scenario, neither beam selection model performed well with LIDAR data. However, Salehi's neural network \ac{NN} showed performance slightly closer to the state-of-the-art. Despite this, the tracking model achieved an impressive 97\% accuracy in top-1. The difference in performance between scenarios t001 and t002 suggests that the analyzed selection \ac{NN} architecture might be site-specific and was not able to handle the non-linear trajectory in the roundabout scenario. This aligns with the findings in \cite{salehi2022deep}, where the \ac{NN} was trained and tested within the same scenario from t001. This indicates that Salehi's architecture might be highly tailored to a specific type of scenario and struggles to generalize to different environments.

In contrast, the proposed \ac{NN} demonstrated strong adaptability to both scenarios, achieving high accuracy in each and even showing improved performance in t002. This underlines the robustness and generability of the proposed architecture in handling diverse scenarios.

\section{Conclusions}
\label{sec:conclusions}
In this paper, we evaluated the performance of \ac{ML} models when using information from LIDAR sensors and GNSS to perform beam tracking in a vehicular B5G scenario.

To process the data fusing  LIDAR and GNSS, we proposed a \ac{CNN}-based model that outperformed the other models. 
This model uses information on the previously chosen beam indices to improve the tracking performance.

Another contribution of this work is the development of a new and public\footnote{https://github.com/AiltonOliveir/AI-Enhanced-MIMO-BeamTracking}  dataset with wireless channels having adequate consistency over time.
This new dataset targets evaluations of beam selection and, especially, beam tracking.

Using this new dataset, the proposed models achieved 84\%  and 97\% of top-1 accuracy for the t001 and t002 scenarios, respectively.

Future works will adopt a \emph{data-centric} approach to the beam tracking problem, aiming at  increasing the size and diversity of the public datasets to promote innovation and systematic and sensible evaluations.


\bibliographystyle{IEEEtran}
\bibliography{main}
\end{document}